\documentclass[seceq]{ptptex}

\usepackage{graphicx}
\usepackage{hyperref}
\usepackage{mciteplus}

\newcommand{\ktt}{k_\perp}
\newcommand{\ptt}{p_\perp}
\newcommand{\qtt}{q_\perp}

\def\p{{\boldsymbol p}}
\def\q{{\boldsymbol q}}

\newcommand{\qs}{Q_\mathrm{s}}
\newcommand{\lqcd}{\Lambda_{_{\rm QCD}}}
\newcommand{\as}{{\alpha_\mathrm{s}}}

\newcommand{\xt}{\boldsymbol{x}_\perp}
\newcommand{\yt}{\boldsymbol{y}_\perp}
\newcommand{\pt}{{\boldsymbol{p}_\perp}}
\newcommand{\qt}{\boldsymbol{q}_\perp}
\newcommand{\kt}{\boldsymbol{k}_\perp}
\newcommand{\nc}{{N_\mathrm{c}}}

\newcommand{\da}{{d_\mathrm{A}}}

\newcommand{\gev}{\textrm{ GeV}}

\newcommand{\nr}[1]{(\ref{#1})} 
\newcommand{\ud}{\mathrm{d}}
\newcommand{\fig}{fig.~}

\newcommand{\eq}{eq.~}

\title{%        %You can use \\ for explicit line-break.
Gluon correlations in the glasma%
}

\author{%       %Use \scshape for the family name.
T. \textsc{Lappi}%
}

\inst{%     %Affiliation, neglected when [addenda] or [errata].
Department of Physics, %
 P.O. Box 35, 40014 University of Jyv\"askyl\"a, Finland and \\
Helsinki Institute of Physics, P.O. Box 64, 00014 University of Helsinki, Finland%
}

\abst{%         %This abstract is neglected when [addenda] or [errata].
The physics of the initial conditions of heavy ion collisions is
dominated by the nonlinear gluonic interactions of QCD. These lead to
the concepts of parton saturation and the
Color Glass Condensate (CGC). We discuss recent progress
in calculating multi-gluon correlations in this framework, prompted
by the observation that these correlations are in fact easier to compute in a
dense system (nucleus-nucleus) than a dilute one (proton-proton).
}

\begin{document}

\maketitle

\section{Introduction}

Bulk particle production in relativistic collisions around midrapidity 
originates from
small $x$ degrees of freedom, predominantly gluons,
 in the wavefunctions of the colliding hadrons or nuclei.
 Because of the $\ln 1/x$ enhancement of soft gluon 
bremsstrahlung this is a dense gluonic system. When the occupation 
numbers of gluonic states in the wavefunction become large enough,
of the order of $1/\as$,
the nonlinear interaction part of the 
Yang-Mills Lagrangian becomes of the same order of magnitude as
the free part. 
This leads to the concept of a transverse 
momentum scale $\qs$, the \emph{saturation scale}, below which the system is
dominated by nonlinear interactions. When the collision energy is high enough
($x$ small enough), $\qs \gg \lqcd$ and the coupling is weak: we are
faced with a \emph{nonperturbative} strongly interacting system
with a \emph{weak coupling constant}. On the other hand, the large occupation 
numbers mean that the system should behave as a \emph{classical} field.
 This suggests a way of organizing
calculations that differs from traditional perturbation theory. Instead 
of developing as a series of powers in $g A_\mu$ one wants to calculate the
classical background field $A^\mu_{\textrm{cl.}}$ and loop corrections 
(which are suppressed by powers of $g$) to all orders in 
$g A^\mu_{\textrm{cl.}}$. The classical gluon field will then be radiated
by the large $x$ degrees of freedom, which are treated as 
effective classical color charges. 
 They 
can be described as random color charges drawn from
a classical probability distribution $W_y[\rho]$ that 
depends on the rapidity cutoff $y=\ln 1/x$ separating the
large and small $x$ degrees of freedom. The dependence of
$W_y[\rho]$ on  $y$ is described by a Wilsonian 
renormalization group equation known by the acronym JIMWLK.
This picture of the 
high energy wavefunction is referred to as the 
Color Glass condensate (CGC, for reviews see 
e.g.~\cite{Iancu:2003xm,*Weigert:2005us,*Gelis:2010nm,*Lappi:2010ek}).

The role of  $W_y[\rho]$ is analoguous to the conventional 
parton distribution function; it is a nonperturbative quantity
whose dependence on one of the kinematical variables of the 
process is described by a weak coupling renormalization group equation.
In the case of pdf's
the appropriate degrees of freedom are individual partons
with a definite momentum, whereas in the case of the CGC 
they are color charges resulting
from interactions of many partons. The distributions
$W_y[\rho]$ are, like pdf's, not (complex) wavefunctions but real
and can be interpreted as probability distributions. 
This is guaranteed by 
\emph{factorization theorems}\cite{Gelis:2008rw,*Gelis:2008ad}.
at large $x$ and the process one is studying.
Factorization can be understood as a statement that 
one has found the right set of degrees of freedom 
in which one can compute physical observables from only the diagonal
elements of the density matrix of the incoming nuclei.

The term glasma~\cite{Lappi:2006fp}  refers to 
the coherent, classical field configuration resulting from the collision of 
two objects described in the CGC framework. The glasma fields are initially longitudinal, 
whence the ``glasma flux tube''~\cite{Dumitru:2008wn,Gavin:2008ev} picture. More importantly
for computing multigluon correlations, they are boost invariant
(to leading order in the QCD coupling) and depend on the 
transverse coordinate with a characteristic correlation 
length $1/\qs$. There are several signals in the RHIC 
data~\cite{Putschke:2007mi,*Daugherity:2008su,*Wenger:2008ts,Abelev:2009dq} that point to strong 
correlations originating from the initial stage of the collision. 
The glasma fields provide a natural framework for understanding
these effects, although much work is still left to do in understanding
the interplay with purely geometrical effects from the fluctuating
positions of the nucleons in the colliding 
nuclei~\cite{Lappi:2009vb}.

 We shall first make some general observations on computing
 multigluon correlations in the glasma, arguing in 
 Sec.~\ref{sec:multig} that they are in some sense simpler
 to compute in a collision of two dense, saturated nuclear wavefunctions
 than in the dilute limit. 
 We shall then, in Sec.~\ref{sec:negbin}
 discuss one application of these ideas to computing
 the multiplicity distribution of gluons in the collision before
 moving to the leading $\ln 1/x$ rapidity dependence of the correlation
 in Sec.~\ref{sec:rapdep}. The application of these ideas to understanding
 correlations observed in the experimental data are discussed
 in Sec.~\ref{sec:ridge}.

\begin{figure}[tb]
\centerline{\resizebox{0.8\textwidth}{!}{
\includegraphics[height=5cm]{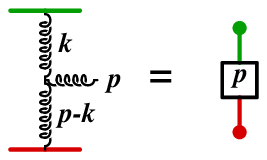}
\rule{5cm}{0pt}
\includegraphics[height=5cm]{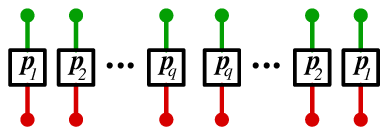}
}}
\caption{Left: Building block, Lipatov vertex coupled to two sources.
Right: combinatorics of the sources. The combinatorial problem is
to connect the dots on the upper and lower side (left- and right moving 
sources) pairwise.
}\label{fig:lipatov} \label{fig:combinat}
\end{figure}

\section{Multigluon correlations in the glasma}
\label{sec:multig}

The gluon fields in the glasma are nonperturbatively strong, $A_\mu \sim 1/g$. 
This means that the gluon multiplicity  is $N \sim 1/\as$. For
a fixed configuration of the classical color sources it is well known that
the multiplicity distribution of produced gluons is Poissonian, 
i.e. $\langle N^2 \rangle - \langle N \rangle^2 = \langle N \rangle$. 
In this case the correlations and fluctuations in the gluon multiplicity
are all quantum effects that appear only starting from the one-loop level, 
i.e. suppressed by a power of the coupling constant $\as$.
The computation in the CGC framework does not end here, however. To calculate
the moments of the gluon multiplicity distribution one must first 
calculate the gluon spectra for fixed configuration of the color charges 
$\rho$ and then average over the probability distribution
$W_y[\rho(\xt)]$.
For the $n$th moment of the multiplicity distribution, i.e. an $n$-gluon correlation,
the leading order result is
\begin{equation}
\left\langle
\frac{\ud N}{\ud^3\p_1}
\cdots 
\frac{\ud N}{\ud^3\p_n}
\right\rangle 
= \left[ \,
\int\limits_{[\rho]} 
W\big[\rho_1(y)\big]  W\big[\rho_2(y)\big]
\left.  \frac{\ud N}{\ud^3\p_1}\right|_{_{\rm LO}}
\cdots
\left.  \frac{\ud N}{\ud^3\p_n}\right|_{_{\rm LO}}
\right],
\end{equation}
where the subsrcipt ``LO'' refers to the single gluon spectrum evaluated 
from the classical field configuration corresponding to a fixed
configuration of color charges.
This averaging, even after the subsequent subtraction of the
appropriate disconnected contributions, introduces a correlation 
already at the leading order in $\as$, i.e. enhanced by 
an additional $1/\as$ compared to the quantum correlations.
A natural example is the negative binomial distribution that 
we shall discuss below, whose variance is 
$\langle N^2 \rangle - \langle N \rangle^2 = 
\langle N \rangle^2/k + \langle N \rangle$. 
One must emphasize here that although these contributions arise
as formally classical correlations in the effective theory that is the
CGC, they are physically also quantum effects, where the weak coupling is 
compensated by a large logarithm of the energy that has been
resummed into the probability distribution $W_y[\rho(\xt)]$. In this
sense the leading correlations are present already in the wavefunctions
of the colliding objects.

\section{Multiplicity distribution}
\label{sec:negbin}

We can then apply this formalism to the calculation of the probability 
distribution of the number of gluons in the glasma~\cite{Gelis:2009wh}.
We shall assume the ``AA'' power counting of sources $\rho$ that
are parametrically strong in $g$, but nevertheless work to the lowest 
nontrivial order in $\rho$. Formally this would correspond
to a power counting $\rho \sim g^{\varepsilon -1}$ with a small 
$\epsilon > 0$. In this limit, as we have discussed, the dominant
contributions to multiparticle correlations come from diagrams that
are disconnected for fixed sources and become connected only
after averaging over the color charge configurations. 
The corresponding two gluon correlation
function was computed in Ref.~\cite{Dumitru:2008wn} and
 generalized to a three gluons in Ref.~\cite{Dusling:2009ar}.
We shall here sketch the derivation~\cite{Gelis:2009wh} 
of the general $n$-gluon  correlation in this simplified limit.

\begin{figure}[tb]
\begin{center}
\resizebox{0.9\textwidth}{!}{
\includegraphics[width=0.35\textwidth]{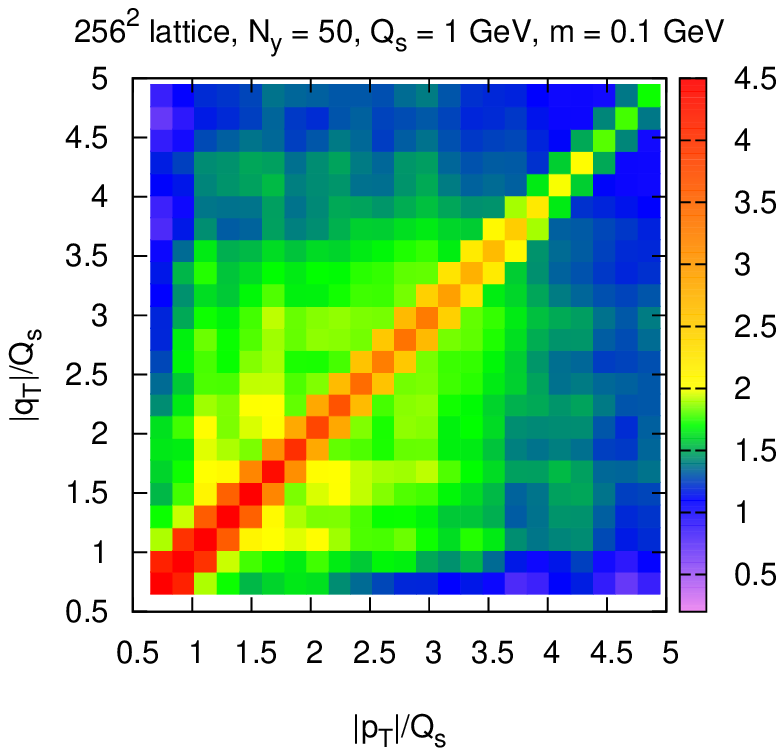}
\rule{1cm}{0pt}
\includegraphics[width=0.35\textwidth]{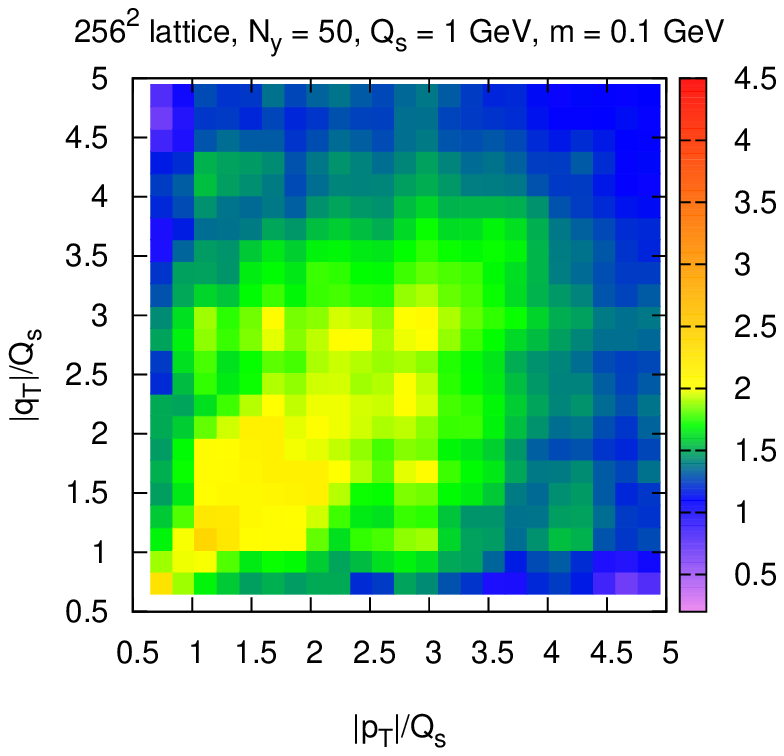}
}
\end{center}
\caption{Numerical evaluation of the two-gluon correlation strength in 
the MV model as a function of the two transverse momenta
$\ptt$ and $\qtt$. Left: the correlation strength
on the ``near side'' $|\varphi_p - \varphi_q| < \pi/2$, right:
the ``away side'' $|\varphi_p - \varphi_q| > \pi/2$.
}\label{fig:kappa2}
\end{figure}

Working with the MV model Gaussian probability distribution
\begin{equation}
W[\rho] = \exp \left[ -\int \ud^2\xt \frac{\rho^a(\xt)\rho^a(\xt)}{g^4\mu^2} \right]
\end{equation}
computing the correlations and the multiplicity distribution
in the linearized approximation
 is a simple combinatorial problem.
Each gluon is produced from two Lipatov vertices 
(see \fig\ref{fig:lipatov} left), 
one in the amplitude and the other in
the complex conjugate. The combinatorial factor is obtained by counting
the different ways of contracting the sources pairwise 
(see \fig\ref{fig:combinat} right).
The dominant contributions are the ones that have, in the dilute limit,
 the strongest infrared divergence 
which is regulated 
by the transverse correlation scale of the problem, $\qs$.
When integrated over the momenta of the produced gluons one 
obtains the factorial moments of the multiplicity, 
which define the whole probability distribution.
 It can be expressed in 
terms of two parameters, the mean multiplicity $\bar{n}$, and a
parameter $k$ describing the width of the distribution.
The result of the combinatorial  exercise is that the 
$q$th factorial moment $m_q$ 
(defined as $\langle N^q \rangle$ minus the corresponding 
disconnected contributions) is
\begin{equation}
 m_q = (q-1)! \,  k \left(\frac{\bar{n}}{k} \right)^q \quad \textrm{with}\quad
k \approx  \frac{  (\nc^2-1)   \qs^2 S_\perp  }{2\pi}
\quad\textrm{and}\quad
\bar{n} =  f_N \frac{1}{\as} \qs^2 S_\perp.
\end{equation}
Here $S_\perp$ is the transverse area of the system.
These moments define a \emph{negative binomial} distribution (NBD) with parameters  
$k$ and $\bar{n}$. The NBD has been known
as a purely phenomenological observation in high energy hadron and nuclear collisions
already for a long  time. 
Also the numerical magnitude of the parameter $k$ obtained from the saturation
scale agrees very well with both pp and AA 
data~\cite{Alner:1985zc,*Ansorge:1988fg,Adare:2008ns}.

In terms of the glasma flux tube picture this result has a natural  interpretation.
The transverse area of a typical flux tube is $1/\qs^2$, and thus there are 
$\qs^2 S_\perp= N_{\textnormal{FT}}$  independent ones. Each of these radiates
particles independently into $\nc^2-1$ color states in a Bose-Einstein distribution
(see e.g.~\cite{Fukushima:2009er}). A sum of $k \approx N_{\textnormal{FT}} (\nc^2-1)$ 
independent Bose-Einstein-distributions is precisely 
equivalent to  a negative binomial 
distribution with parameter $k$.

\begin{figure}[tb]
\centerline{
\resizebox{0.45\textwidth}{!}{
\includegraphics[height=5cm]{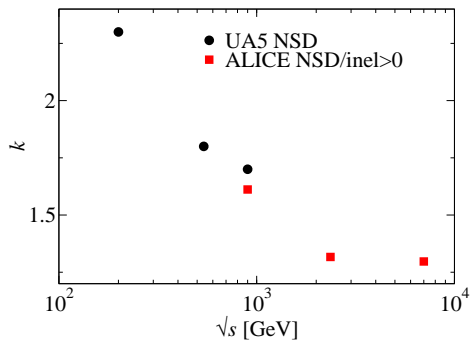}
}
}
\caption{
The parameter $k$  in pp collisions measured by the UA5~\cite{Alner:1985zc},
and ALICE~\cite{Aamodt:2010ft,*Aamodt:2010pp}. For UA5 the NBD fit is done by 
the experiment. The fit to ALICE data is done by us and does not take into account the
correlated errors in the experimental data points.
}\label{fig:nbd}
\end{figure}

This calculation predicts that the $k$ parameter should increase with energy
as $k\sim \qs^2 \sim \sqrt{s}^\lambda$. This is ideed true in the heavy ion data; 
$k$ is reported to  increase from $\sqrt{s}=A 62\gev$ to  $A 200\gev$
in by the PHENIX collaboration~\cite{Adare:2008ns}.
However, the interpretation of the heavy ion data is 
complicated by the purely geometrical fluctuations from the different
impact parameters probed in one centrality bin.

The proton-proton collision system is much smaller and fluctuations at lower energies
are still mostly dominated by the dilute edge of the collision system, which 
has a  Poissonian nature (i.e. $k\to \infty$). Our
calculation formally assumes $\qs^2 S_\perp \gg 1$, and we expect the 
growing behavior of $k$ with energy to eventually take over at high enough energy.
Since the fluctuations are more dominated by the edge region than the mean multiplicity,
it is natural to expect this transition to a genuine high energy regime
to be visible later in the fluctuations than it is in the average. At
LHC energies (see the data in Ref.~\cite{Aamodt:2010ft,*Aamodt:2010pp}) 
the transition to an increasing $k$ seems to be starting
as shown in \fig\ref{fig:nbd}.

\section{Rapidity dependence}

\label{sec:rapdep}

The general discussion of Sec.~\ref{sec:multig} on the different nature of 
multigluon correlations in the ``AA'' case applies also to the rapidity dependence.
Until now we have only been discussing gluon production in a rapidity interval smaller
than $1/\as$. For this we needed only the correlations between the color charges
$\rho(\xt)$ measured at this same rapidity. To understand the rapidity dependence
of the correlations one needs also the correlation between color charges at
different rapidities, $\langle \rho_y(\xt) \rho_{y'}(\yt) \rangle$. 
Also this information is contained in the JIMWLK renormalization group evolution,
at least to  leading $\ln 1/x$ accuracy~\cite{Gelis:2008sz,*Lappi:2009fq}.
An intuitive description of the resulting correlations is provided by the formulation
of JIMWLK as a Langevin equation in the space of Wilson lines formed from the color 
charges. In this picture the evolution proceeds in individual trajectories
along an increasing rapidity.

A first attempt of a realistic estimate of the rapidity dependence 
of two-gluon correlations is performed in Ref.~\cite{Dusling:2009ni}. Evaluating 
the two gluon correlation in a dilute limit in a $\ktt$-factorized 
approximation, but keeping the general  structure resulting from the JIMWLK evolution
leads to the following expression:
\begin{multline}
\label{eq:double-inclusive-4}
C(\p,\q)
=
\frac{\as^{2}}{16 \pi^{10}}
\frac{N_c^2(N_c^2-1) S_\perp}{\da^4\; \pt^2\qt^2}
\\
\times
\bigg\{
\int \ud^2\kt
\Phi_{A_1}^2(y_p,\kt)\Phi_{A_2}(y_p,\pt-\kt)
\Big[
\Phi_{A_2}(y_q,\qt+\kt)
+
\Phi_{A_2}(y_q,\qt-\kt)
\Big]
\\
+
\Phi_{A_2}^2(y_q, \kt)\Phi_{A_1}(y_p,\pt-\kt)
\Big[
\Phi_{A_1}(y_q,\qt+\kt)
+
\Phi_{A_1}(y_q,\qt-\kt)
\Big]
\bigg\}.
\end{multline}
Note the very different structure of this correlation compared to one where
the gluons would be produced from the same diagram for fixed sources. The two 
gluon correlation function is proportional to the product of
\emph{four} unintegrated gluon distributions, with three of them evaluated
at the rapidity of one of the produced gluons and only one at the other. This 
structure is a direct consequence of the nature of JIMWLK evolution.

\begin{figure}[tb]
\centerline{
\resizebox{0.9\textwidth}{!}{
\includegraphics[height=5cm]{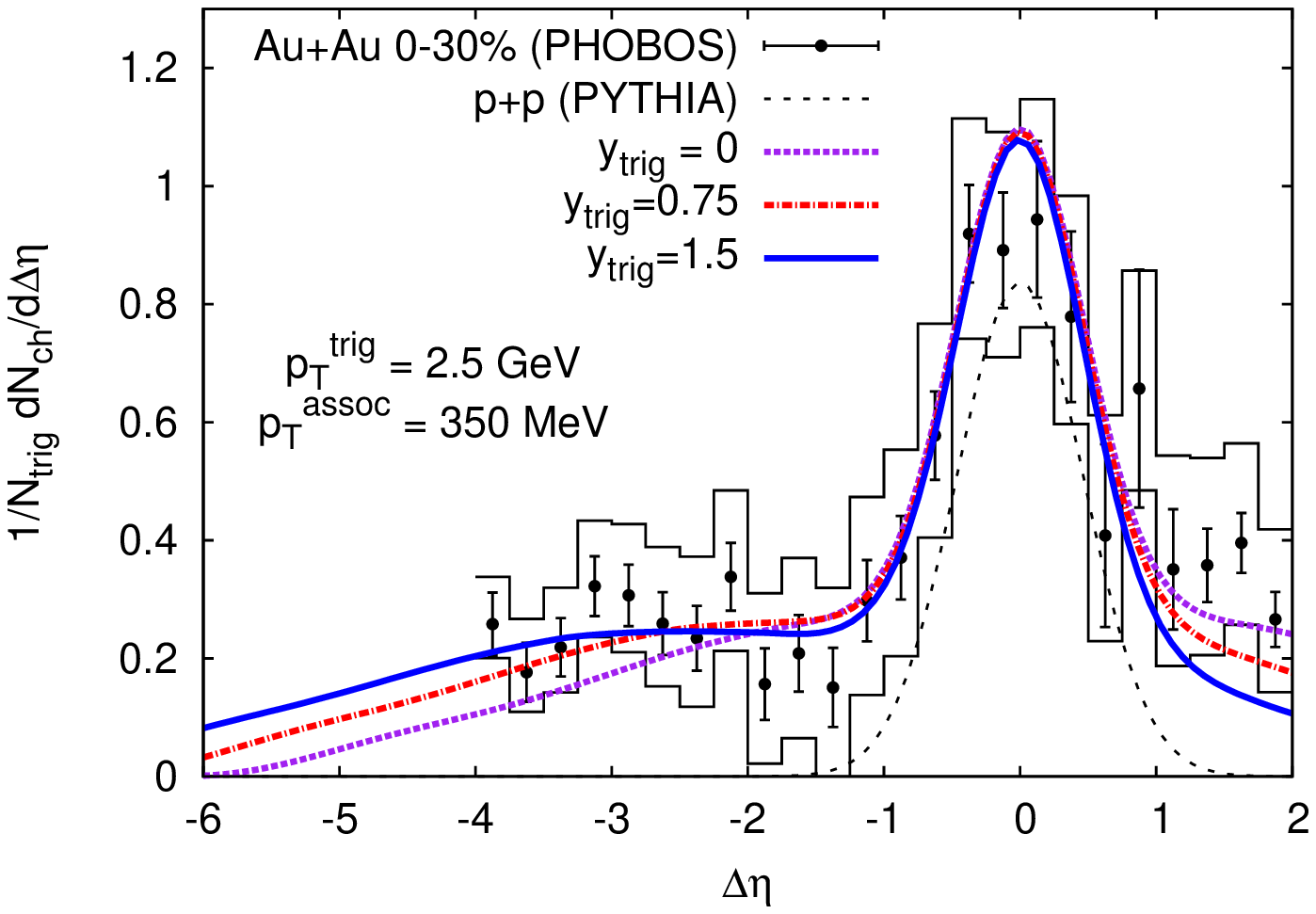}
\includegraphics[height=5.222cm]{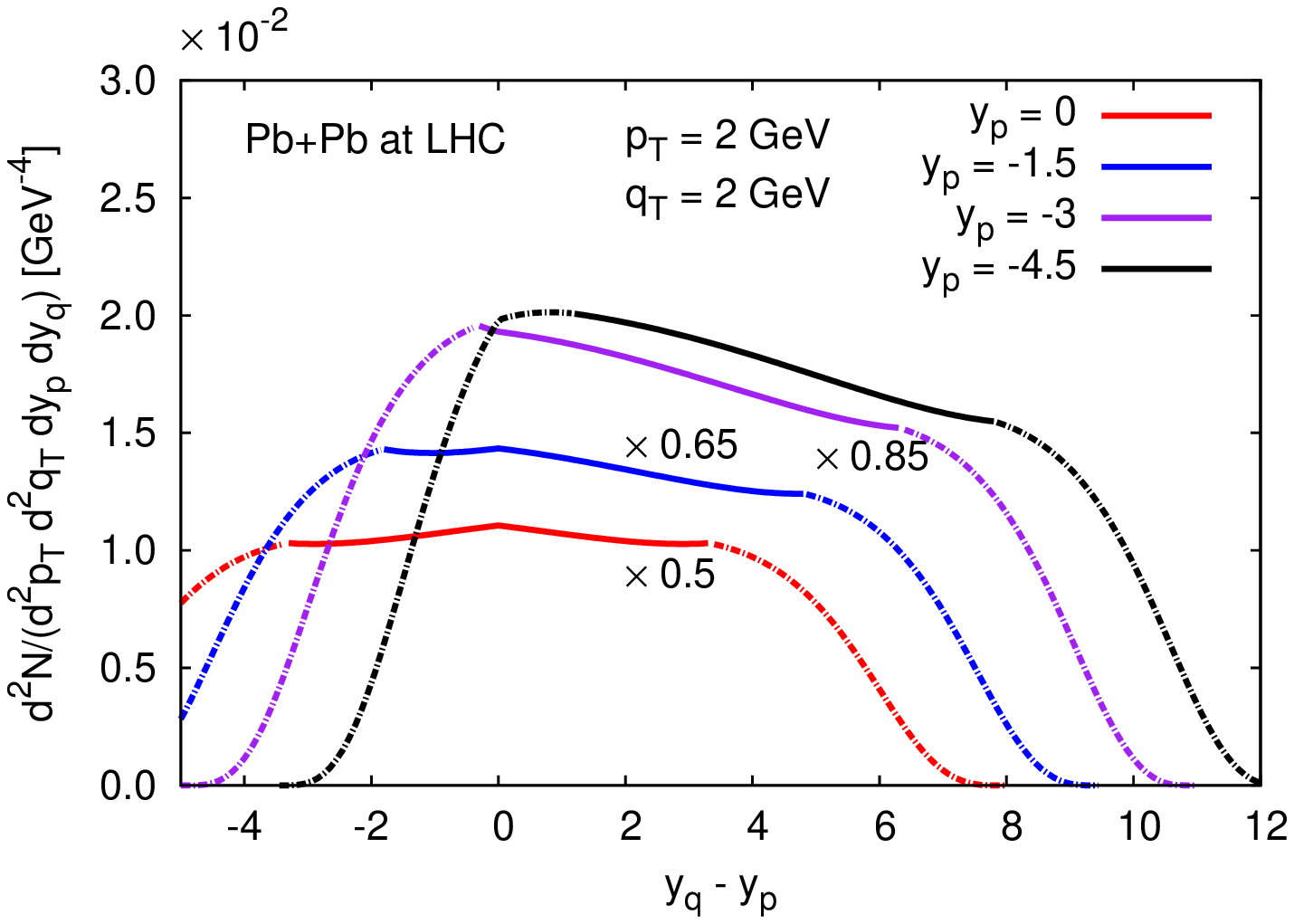}
}
}
\caption{
Left: Comparison of a two-particle correlation computed using 
\eq\nr{eq:double-inclusive-4} supplemented with a short-range correlation 
contribution from PYTHIA with PHOBOS data.
Right: Rapidity correlation at LHC energies $\ktt$-factorization approximation.
}\label{fig:rapidity}
\end{figure}

\section{The ridge in nuclear and proton collisions}
\label{sec:ridge}

 The ``ridge'' is a feature on the ``near side'' 
of the two particle correlation, around $\Delta \phi \approx 0$
azimuthal separation between the two particles. It
extends up to high values of $\Delta \eta,$ the pseudorapidity
difference between the two particles.
This feature was first  seen in nucleus--nucleus
collisions at
RHIC~\cite{Adams:2004pa,Adams:2005ph,Adare:2008cq,Alver:2009id,Alver:2010rt}.
The ridge
was seen in high multiplicity (central) events in nucleus-nucleus
collisions in Cu+Cu collisions at $\sqrt{s}=62.4$~GeV and in Au+Au
collisions at $\sqrt{s}=200$~GeV. The STAR detector observed 
this correlation for both $p_\perp$-triggered~\cite{Adams:2005ph} and
untriggered~\cite{Adams:2004pa} pair correlations in the whole STAR TPC
acceptance of $\Delta \eta \leq 2$. 
The PHOBOS experiment~\cite{Alver:2009id} observed the $p_\perp$-triggered 
correlation at much larger rapidity separations,
initially up to $\Delta \eta\sim 4$, extended more
recently~\cite{Alver:2010rt} to $\Delta\eta \sim 5$. 
The long range correlation
structure disappears for lower multiplicity peripheral events in 
nucleus-nucleus collisions and are also absent in deuteron-gold and
proton-proton ``control'' experiments at the same energies.

The ``standard'' glasma explanation~\cite{Dumitru:2008wn} 
of the azimuthal structure of the 
ridge relies on a collimation effect from radial flow combined
with the long range rapidity correlation from the boost invariant 
color fields.
The correlation computed from \eq\nr{eq:double-inclusive-4} 
is compared to PHOBOS data in \fig\ref{fig:rapidity}.
The $\ktt$-factorized approximation gives a very inaccurate description of the 
gluon spectrum in the transverse momentum regime $\ptt \sim \qs$ where
the bulk of the particles are produced~\cite{Blaizot:2010kh}. 
Equation~\nr{eq:double-inclusive-4} has also been derived in the 
approximation, true only in the linearized case,
that the unequal rapidity correlation of two color charge densities 
is equal to the unintegrated gluon distribution at the smaller one of 
these rapidities. As of yet there is no calculation of how much this 
approximation is violated in the full JIMWLK evolution
(see, however, \cite{Dumitru:2010mv}). 
The results presented
in \fig\ref{fig:rapidity} are therefore not the final word on the subject, 
alhough it is reassuring that such a simple approximation seems to agree 
rather well with the experimental result.
 A numerical evaluation~\cite{Lappi:2009xa} 
of the second moment of the 
distribution, parametrized in terms of
\begin{equation}
\kappa_2(\pt,\qt) = \qs^2 S_\perp \left( 
\frac{\ud^2 N}{\ud^2\pt \ud^2\qt}
-
\frac{\ud N}{\ud^2\pt } \frac{\ud N}{\ud^2\qt }
\right) \bigg/ \frac{\ud N}{\ud^2\pt } \frac{\ud N}{\ud^2\qt }
\end{equation}
is shown in Fig.~\ref{fig:kappa2}. It confirms the expectations 
of Ref.~\cite{Gelis:2009wh} on the magnitude of the correlation. 
In contrast to the approximation made in  Ref.~\cite{Gelis:2009wh} there is, 
however, some dependence on the momenta $\pt,\qt$, which is crucial
for the discussion of the correlation in smaller systems where transverse
flow is smaller or absent.

\begin{figure}[tb]
\begin{center}
\includegraphics[width=0.45\textwidth]{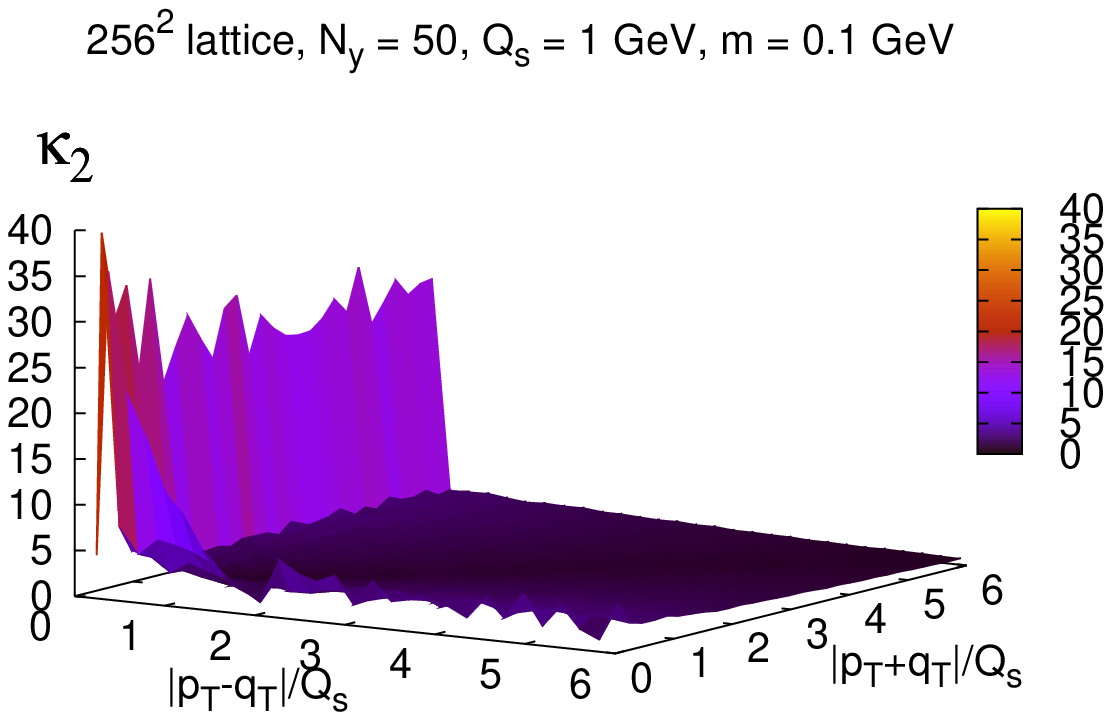}
\includegraphics[width=0.45\textwidth]{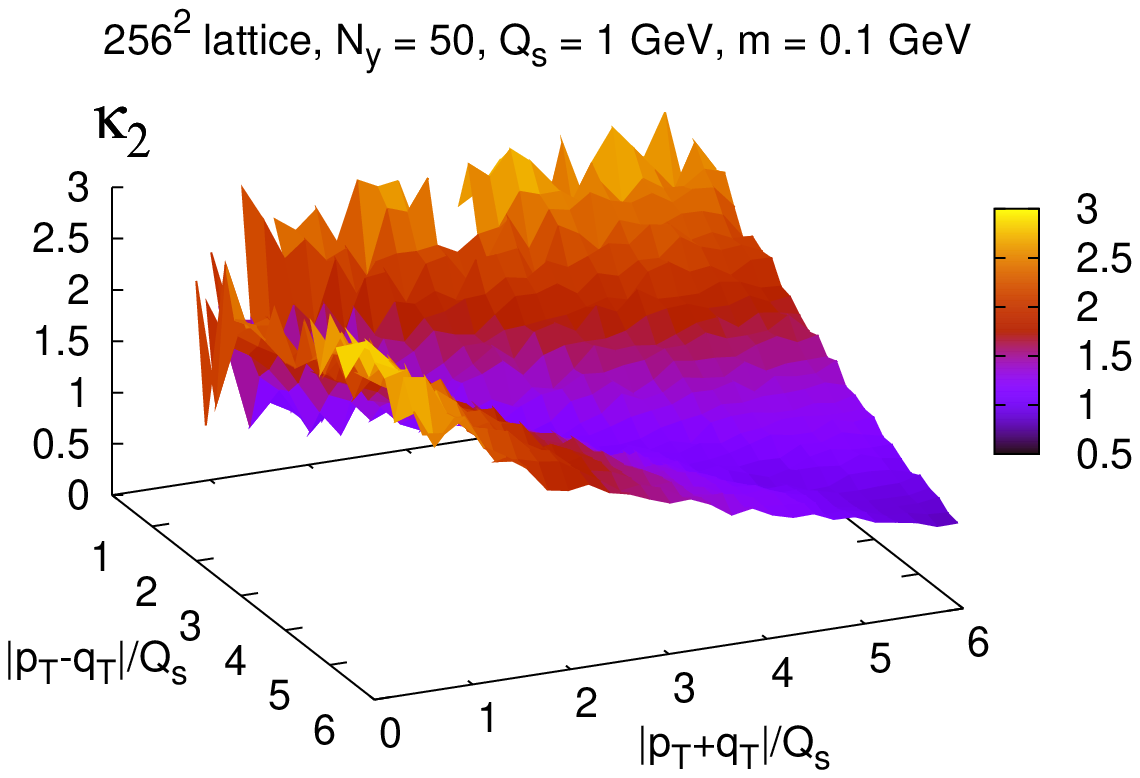}
\end{center}
\caption{Two-gluon correlation strength in 
the MV model. Both figures show the same data; the large delta function
peaks have been removed from the one on the right to illustrate the structure in 
the rest of phase space.}\label{fig:kappavsdiff}
\end{figure}
 
More recently the CMS collaboration has reported the observation of a ``ridge''
structure also in proton-proton collisions~\cite{Collaboration:2010gv}.
 The ridge is
seen only in the moderate $p_\perp,q_\perp$ range and systematically
vanishes for $p_\perp,q_\perp \lesssim 1$~GeV and 
$p_\perp,q_\perp \gtrsim 3$~GeV. 
Although the explanation in terms of the collimating effect of the transverse
flow is most likely present to some degree in high multiplicity proton-proton 
collisions at LHC energies, this observation points to a ridge-like effect
being present already in the initial scattering process. This is indeed the 
picture resulting from the correlations in the glasma. 

In terms of the $\ktt$-factorized assumption~\nr{eq:double-inclusive-4}
the CMS data was recently discussed in Ref.~\cite{Dumitru:2010iy}.
The dependence of the CMS ridge on transverse momentum was found
to qualitatively agree with the CMS result.
In the full nonperturbative calculation this same collimation
effect already in the initial stage is visible in \fig\ref{fig:kappavsdiff},
where the correlation is plotted as a function of 
$|\pt-\qt|$ and $|\pt+\qt|$. The production of gluons from a coherent
classical field results in an enhanced near side correlation for momenta
$\pt$ and $\q$ that are close to parallel, even in the absence of
transverse flow.

\section{Conclusion}
Most experimental observables do not probe the glasma
initial state of directly, because the system goes through a complicated
time evolution before the hadronization stage. A good candidate for an 
 experimental probe giving direct access to the initial state
is provided by different kinds of
correlation measurements. These have indeed been a focus of 
both experimental and theoretical activity recently. 
We have argued here that the glasma picture of the initial stages
of a heavy ion collision is the natural framework
to understand the origin of these correlations.

\section*{Acknowledgements}
The author thanks the Yukawa Institute of Theoretical Physics 
t Kyoto University for hospitality
and support during YIPQS Workshop ``High energy strong interactions 2010.''
The author is supported by the Academy of Finland, contract 126604.

\bibliographystyle{h-physrev4mod2M}
\bibliography{spires}

\end{document}